\documentclass[aps,prb,preprint,showpacs]{revtex4}
\usepackage{graphicx}
\usepackage{helvet}
\usepackage{amsmath,amssymb}
\usepackage{bm}

\begin{document}

\title{Electron Correlations in the Quasi-Two-Dimensional Organic Conductor 
$\theta$-(BEDT-TTF)$_{2}$I$_{3}$ investigated by $^{13}$C NMR}

\author{Michihiro Hirata$^{1}$, Kazuya Miyagawa$^{1}$, Kazushi Kanoda$^{1}$, 
Masafumi Tamura$^{2}$}
	
\affiliation{
$^{1}$ Department of Applied Physics, University of Tokyo, Bunkyo-ku, Tokyo, 113-8656, Japan \\
$^{2}$ Department of Physics, Faculty of Science and Technology, Tokyo University of Science, 
Noda, Chiba, 278-8510, Japan \\}

\date{\today}

\begin{abstract}
	We report a $^{13}$C-NMR study on the ambient-pressure metallic phase of the layered organic conductor 
$\theta$-(BEDT-TTF)$_{2}$I$_{3}$ [BEDT-TTF: bisethylenedithio-tetrathiafulvalene],
which is expected to connect the physics of correlated electrons and Dirac electrons under pressure.  
	The orientation dependence of the NMR spectra shows that all BEDT-TTF molecules 
in the unit cell are to be seen equivalent from a microscopic point of view. 
	This feature is consistent with the orthorhombic symmetry of the BEDT-TTF sublattice and also 
indicates that the monoclinic $I_{3}$ sublattice, which should make three molecules in the unit cell nonequivalent, 
is not practically influential on the electronic state in the conducting BEDT-TTF layers at ambient pressure.  
	There is no signature of charge disproportionation in opposition to most of the $\theta$-type BEDT-TTF salts. 
	The analyses of NMR Knight shift, $K$, and the nuclear spin-lattice relaxation rate, $1/T_{1}$, 
revealed that the degree of electron correlation, evaluated by the Korringa ratio [$\varpropto 1/(T_{1}TK^{2}$)], 
is in an intermediate regime. 
	However, NMR relaxation rate $1/T_{1}$ is enhanced above $\sim$ 200K, which possibly indicates 
that the system enters into a quantum critical regime of charge-order fluctuations as suggested 
theoretically. 
\end{abstract}



\maketitle

\section{Introduction}
\label{intro}
\begin{figure}
 \includegraphics[width=8.6cm]{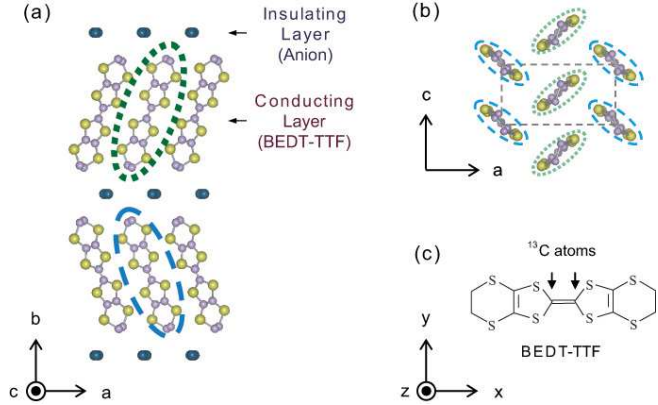}
  \caption{\label{Fig_1} (Color online) (a) Side view of the orthorhombic structure 
of $\theta$-I$_{3}$.\cite{Ref_08_Kobayashi, Ref_33_Kobayashi} 
(b) Top view of the schematic arrangements of BEDT-TTF molecules in the conducting $ac$ plane. 
(c) Molecular principal axes of a BEDT-TTF molecule. 
Arrows indicate the positions of $^{13}$C atoms. 
In Figures (a) and (b), two molecules, represented by 
dotted and dashed ellipses, become unequal in their geometric arrangements 
with respect to the field directions of $H$//$bc$ and $H$//$ac$, respectively.}
\end{figure}

	The layered organic compounds (BEDT-TTF)$_{2}$X exhibit fascinating electronic phases with 
superconductivity,\cite{Ref_01_Ishiguro} Mott localization,\cite{Ref_02_Miyagawa} 
spin liquid,\cite{Ref_03_Shimizu} charge ordering,\cite{Ref_04_Mori, Ref_05_Miyagawa} 
massless Dirac fermions,\cite{Ref_06_Tajima, Ref_07_Katayama} and so on,\cite{Ref_Add_1} 
where X is monovalent anion and BEDT-TTF is bisethylenedithio-tetrathiafulvalene. 
	The system is constructed fom alternate stacking of 2D conducting layers of the donor molecule 
BEDT-TTF and insulating layers of anion X [Fig.~\ref{Fig_1}(a)]. 
	The anion extracts an electron from two BEDT-TTF molecules, i.e., (BEDT-TTF)$^{+0.5}$ : 
X$^{-}$, which results in a quarter-filled hole band through inter-molecular overlaps of 
the molecular orbital. 

	The title compound, $\theta$-(BEDT-TTF)$_{2}$I$_{3}$ (abbreviated to $\theta$-I$_{3}$), 
is known as a typical 2D metal with highly symmetric zigzag alignment of BEDT-TTF molecules 
in the conducting layer [Fig.~\ref{Fig_1}(b)]. \cite{Ref_08_Kobayashi}
	It is metallic in the whole temperature range,\cite{Ref_08_Kobayashi}
and the Fermi surface has been confirmed by quantum oscillation 
\cite{Ref_09_Kajita, Ref_10_Klepper, Ref_11_Tokumoto, Ref_12_Tamura, Ref_13_Terashima, 
Ref_14_Salameh} and optical \cite{Ref_15_Tamura, Ref_16_Oshima} experiments. 
	Magnetic susceptibility exhibits weak temperature dependence,\cite{Ref_14_Salameh, Ref_17_Wang} 
characteristic of the Pauli paramagnetism. 
	Moreover, superconductivity is observed at $T_{\textrm{C}}$ of ca. 3.6K. \cite{Ref_08_Kobayashi, Ref_18_Kajita}

	Meanwhile, intense investigations have revealed that other $\theta$-type 
salts with closely associated molecular arrangements, $\theta$-(BEDT-TTF)$_{2}$MZn(SCN)$_{4}$ [M = Rb and Cs] 
(abbreviated to $\theta$-MZn), undergo metal-to-insulator (MI) transitions at 
$T_{\textrm{MI}}$ = 195K ($\theta$-RbZn) and 20K ($\theta$-CsZn), \cite{Ref_04_Mori} 
accompanied by a spontaneous charge ordering (CO) \cite{Ref_05_Miyagawa} and 
a glassy charge disproportionation, \cite{Ref_19_Chiba} respectively. 
	Even in the conducting state above $T_{\textrm{MI}}$, 
where resistivity is temperature insensitive in these materials, \cite{Ref_04_Mori} $^{13}$C-NMR line shows 
a gradual broadening toward $T_{\textrm{MI}}$, 
which is thought of as a manifestation of charge fluctuations. \cite{Ref_05_Miyagawa, Ref_19_Chiba} 
	The overall nature implies that the electron correlations play decisive roles in the electronic phases of $\theta$-MZn. 
	Despite the close similarities of the molecular arrangements in
$\theta$-MZn and $\theta$-I$_{3}$, their contrasting ground states demonstrate a variation 
in the electron correlations and/or bandwidth among the $\theta$-type family, as pointed out by 
Mori \textit{et al}. \cite{Ref_04_Mori} 

	Application of hydrostatic pressures drastically alters the electronic state 
in $\theta$-I$_{3}$. 
	Pressure dependence of the room-temperature resistivity shows a discontinuous jump 
at $\sim$ 0.5GPa. \cite{Ref_20_Tamura} 
	Under pressures above this critical value, $\theta$-I$_{3}$ shows weak temperature dependence 
in resistivity, \cite{Ref_21_Tajima} a large increase in Hall coefficient, \cite{Ref_21_Tajima} 
and $T^{3}$-dependence in nuclear spin-lattice relaxation rate, $1/T_{1}$, below $\sim$ 20K 
\cite{Ref_22_Miyagawa} with decreasing temperature, indicating the presence of a linearly dispersive 
energy band near the Fermi level. \cite{Ref_23_Dora, Ref_24_Katayama} 
	Notice that the same properties are also observed in $\alpha$-(BEDT-TTF)$_{2}$I$_{3}$ 
($\alpha$-I$_{3}$) at high pressures approximately 
above 1.5GPa, \cite{Ref_06_Tajima, Ref_25_Hirata} where a tilted massless Dirac cone is 
predicted by band-structure calculations. \cite{Ref_07_Katayama, Ref_24_Katayama} 
	$\alpha$-I$_{3}$ includes three nonequivalent molecules in the unit cell with a 
zigzag molecular arrangement as in the $\theta$-type materials. \cite{Ref_26_Bender} 
	It is suggested that this $\alpha$-type local site symmetry plays a significant role 
in the realization of the tilted Dirac cone in $\alpha$-I$_{3}$. 
\cite{Ref_07_Katayama, Ref_27_Mori, Ref_28_Asano} 
	Similar argument is also expected in $\theta$-I$_{3}$ under pressures, 
although the crystal structure above $\sim$ 0.5GPa is not yet determined. 

	The lattice symmetry at ambient pressure is also complicated in $\theta$-I$_{3}$; 
	Kobayashi \textit{et al}. \cite{Ref_08_Kobayashi} report a monoclinic structure 
(space group $P2_{1}/c$) for the whole crystal, and an orthorhombic symmetry ($P_{\textrm{nma}}$) for the BEDT-TTF sublattice in terms of x-ray diffraction measurements. 
	Band-structure calculation predicts small Fermi pockets for the monoclinic structure, and a large Fermi surface for the orthorhombic sublattice.\cite{Ref_08_Kobayashi} 
	The quantum oscillation and optical reflectance measurements support the latter.\cite{Ref_09_Kajita, Ref_10_Klepper, Ref_11_Tokumoto, Ref_12_Tamura, Ref_13_Terashima, Ref_14_Salameh, Ref_15_Tamura, Ref_16_Oshima} 
	These facts suggest that the monoclinicity of the crystal, which mainly originates from the arrangements of triiodine (I$_{3}$) molecules,\cite{Ref_08_Kobayashi} seems to make only a negligible influence on the electronic structure.
	To further unravel this point, it is needed to characterize the local electronic state on BEDT-TTF sites microscopically.	

	As seen above, $\theta$-I$_{3}$ is a unique material connecting the physics of 
correlated electrons and Dirac electrons, both of which are among the intensively studied issues 
in condensed matter physics. 
	In the present work, we aim at elucidating the electronic structure in $\theta$-I$_{3}$ 
at ambient pressure by means of $^{13}$C-NMR experiments for the first time. 
	This is expected to provide a basis for understanding the conducting state 
in the proximity of the CO phase and the massless Dirac fermions in organic conductors. 
	First, we measured the orientation dependence of the $^{13}$C-NMR spectra against 
applied magnetic field to evaluate the local-site symmetry and the hyperfine-shift tensor 
at the central $^{13}$C sites in BEDT-TTF molecules [Fig.~\ref{Fig_1}(c)]. 
	Because the hyperfine interaction between electron and $^{13}$C-nuclear spins is 
highly anisotropic in (BEDT-TTF)$_{2}$X compounds, \cite{Ref_29_Kawamoto} 
the hyperfine-shift tensor $\delta = (\delta^{xx}, \delta^{yy}, \delta^{zz})$ at the $^{13}$C sites 
is thereby determined, where $x, y$, and $z$ represent the principal axes of the molecule 
[see Fig.~\ref{Fig_1}(c)]. 
	Then, we measured the Knight shift, $K$, and the nuclear spin-lattice relaxation rate, 
$1/T_{1}$, and determined local spin susceptibility and electron correlations 
in a quantitative manner, using the hyperfine-shift tensor determined. 
	Based on the experimental data and analyses, we discuss the nature of electronic state in 
$\theta$-I$_{3}$ at ambient pressure.

\section{EXPERIMENTAL}
\label{experimental}
	Single crystal of $\theta$-I$_{3}$ was prepared by the conventional electrochemical method. 
	For the $^{13}$C-NMR measurements, the central carbon sites in BEDT-TTF molecules were 
selectively enriched by $^{13}$C atoms 
(nuclear spin $I$ = 1/2, $\gamma_{\textrm{n}}/2\pi$ = 10.7054MHz/T) 
with 99$\%$ concentration [Fig.~\ref{Fig_1}(c)], where a large spin density 
is expected in the highest occupied molecular orbital (HOMO) which is known to give the main 
contribution to the electronic bands at the Fermi level.~\cite{Ref_02_Miyagawa} 
	All NMR measurements were performed for a single crystal at ambient pressure 
in a magnetic field $H$ of approximately 6.00T, which is rotated within the conducting $ac$ plane 
and the $bc$ plane [see Fig.~\ref{Fig_1}]. 
	NMR signals were obtained through the fast Fourier transformation of the so-called solid-echo 
signals, and the $^{13}$C-resonance frequency of tetramethylsilane [(CH$_{3}$)$_{4}$Si, TMS] 
was used as the origin of the NMR shift. 
	The nuclear spin-lattice relaxation rate, $1/T_{1}$, was determined by 
the standard saturation and recovery method. 
	The relaxation curves of nuclear magnetization were well fitted to single exponential 
functions over a decade.

\section{RESULTS AND DISCUSSIONS}
\subsection{NMR spectra, local-site symmetry, and hyperfine-shift tensors}
\label{3a}
\begin{figure}
  \includegraphics[width=8.6cm]{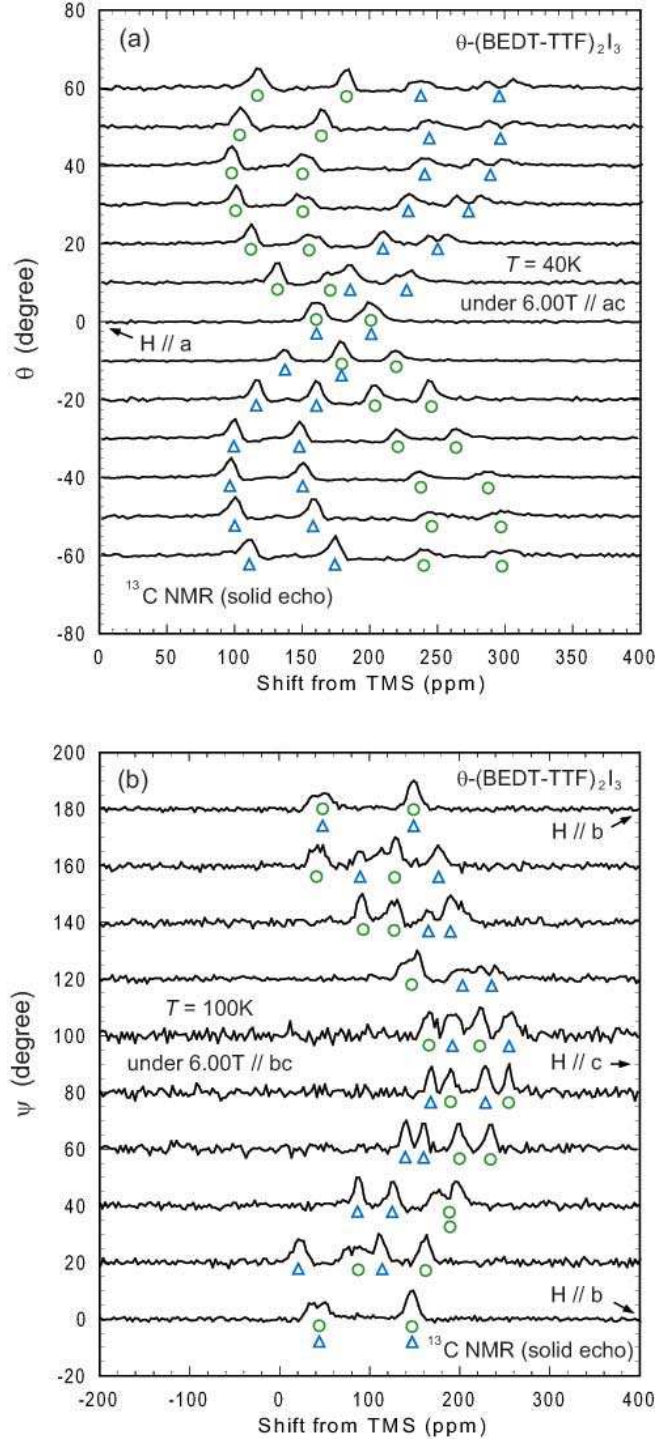}
  \caption{\label{Fig_2} (Color online) Orientation dependence of the $^{13}$C-NMR spectra 
(a) under $H$ within $ac$ plane (40K) and 
(b) $bc$ plane (100K). 
$\theta$ and $\psi$ stand for the field angles measured from $a$ and $b$ axes, respectively. 
Symbols represent different Pake doublets stemming from the two differently oriented molecules 
in the double-decker unit cell (ellipses in Fig.~\ref{Fig_1}).}
\end{figure}

	The local-site symmetry at the central $^{13}$C positions of BEDT-TTF molecules 
[Fig.~\ref{Fig_1}(c)] is examined by the angular profile of $^{13}$C-NMR spectra, 
which were measured with changing 
(i) the angle $\theta$ between the crystal $a$ axis and the applied field $H$ in the $ac$ conducting 
plane [Fig.~\ref{Fig_1}(b)], and 
(ii) the angle $\psi$ between the crystal $b$ axis and $H$ in the $bc$ plane [Fig.~\ref{Fig_1}(a)]. 
Figures \ref{Fig_2}(a) and \ref{Fig_2}(b) show the typical angular dependence of $^{13}$C-NMR 
spectra under $H$ applied within $ac$ plane (at 40K) and $bc$ plane 
(at 100K), respectively. 
	Under all field orientations, the NMR spectra consist approximately of four lines 
at all measured temperatures. 
	Two pairs of lines in Fig.~\ref{Fig_2}(a) exhibit out-of-phase angular dependence 
against $\theta$, whereas an in-phase variation is seen against $\psi$ in Fig.~\ref{Fig_2}(b).

	As we mentioned in Section~\ref{intro}, x-ray diffraction experiments suggest that the crystal structure is monoclinic in $\theta$-I$_{3}$ (with space group $P2_{1}/c$).\cite{Ref_08_Kobayashi}
	In this symmetry, three molecules are crystallographically nonequivalent and locate in each layer of the double-decker unit cell with only two of them possessing inversion symmetry in the molecular center as in $\alpha$-I$_{3}$.\cite{Ref_26_Bender, Ref_32_Hirata, Ref_33_Kobayashi} 
	In the $^{13}$C-NMR spectra, each nonequivalent molecule gives distinct resonance lines. 
	Furthermore, due to the nuclear dipole interaction between the adjacent $^{13}$C nuclei in a molecule [Fig.~\ref{Fig_1}(c)], 
each resonance line from a molecule splits into a doublet or a quartet depending 
on whether the molecule has the inversion symmetry or not. \cite{Ref_29_Kawamoto} 
	In our case, three molecules in a layer (one without and two 
with inversion centers) are unequal against arbitrary field directions. 
	Hence, in total, eight lines (one quartet and two doublets) are expected under the $ac$-plane-field 
geometry ($H$//layer), and they should be doubled under the $bc$-plane condition where the adjacent layers become unequal against magnetic field. 
	The observed NMR spectra show, however, only four lines under all field directions [Figs.~\ref{Fig_2}(a) and \ref{Fig_2}(b)].
	This indicates that the local site differences among molecules are negligibly small in $\theta$-I$_{3}$,
or equivalently the influence of I$_{3}$ alignments, which is responsible for the monoclinicity of the crystal,\cite{Ref_08_Kobayashi} is not significant on the BEDT-TTF layers. 
	This is in line with the x-ray diffraction measurement predicting orthorhombic symmetry for the BEDT-TTF sublattice \cite{Ref_08_Kobayashi} as the observed four-line structure can be well explained by this picture as follows: 
	in the orthorhombic symmetry, all molecules are crystallographically equivalent and the inversion center locates exactly in-between the two adjacent $^{13}$C atoms in a BEDT-TTF. 
	This results in the four-line structure in the NMR spectra (namely, two pairs of doublets) due to 
(i) two differently oriented BEDT-TTFs in a layer [Fig.~\ref{Fig_1}(b)] (in neighboring layers [Fig.~\ref{Fig_1}(a)]) and 
(ii) the nuclear dipole interaction as the field $H$ is applied parallel to the $ac$ ($bc$) plane. 
	(We note that a small peak splitting seen in several spectra, e.g., at 
$\theta$ = 20$^{\circ}$, 30$^{\circ}$, 40$^{\circ}$, 50$^{\circ}$, and 60$^{\circ}$ 
in Fig.~\ref{Fig_2}(a), is attributable to a slight misalignment of the rotating plane of $H$ from 
$ac$ plane.) 

	The situation becomes clearer when we look into the angular dependence of the NMR line structure 
in more detail. 
	Based on the orthorhombic symmetry, the observed spectra can be well assigned to two Pake doublets 
-- circles and triangles in Figs.~\ref{Fig_2}(a) and \ref{Fig_2}(b) --  
stemming from the molecules denoted by dotted and dashed ellipses in Fig.~\ref{Fig_1}(b) [for Fig.~\ref{Fig_2}(a)] and Fig.~\ref{Fig_1}(a) [for Fig.~\ref{Fig_2}(b)]. 

\begin{figure}
  \includegraphics[width=8.6cm, bb=0 0 330 600]{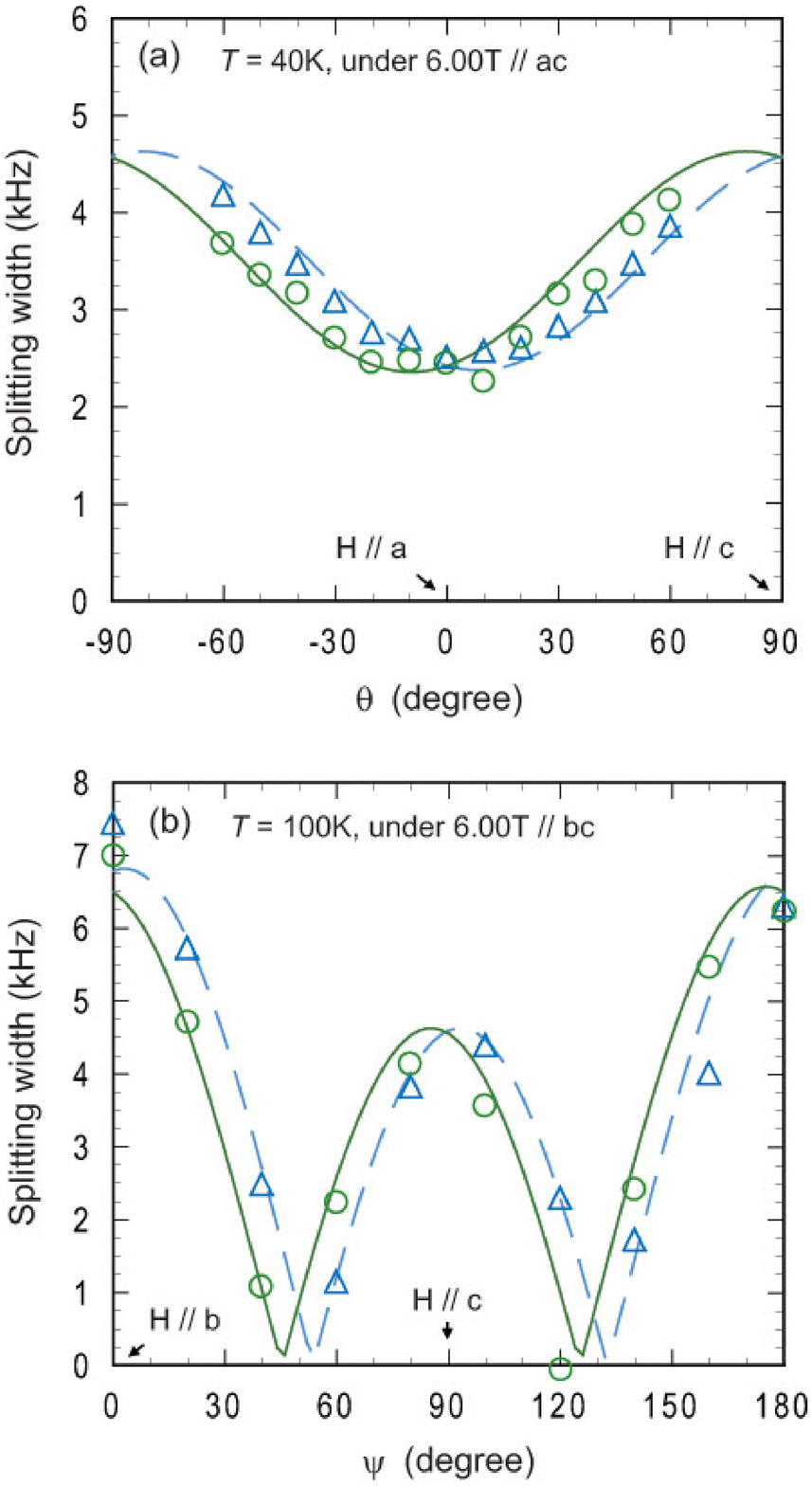}
  \caption{\label{Fig_3} (Color online) Angular dependence of the nuclear dipole splitting 
width $d$ for the two differently oriented molecules in the unit cell (symbols; see Fig.~\ref{Fig_1}) under 
(a) $H//ac$ (40K) and (b) $H//bc$ (100K), which are extracted from Figs.~\ref{Fig_2}(a) and \ref{Fig_2}(b), respectively. 
	The same symbols are used as in Fig.~\ref{Fig_2}. 
	Calculated angular dependence is shown by solid and dashed curves based on the orthorhombic symmetry 
reported by Kobayashi \textit{et al}. \cite{Ref_33_Kobayashi} and the optimal $^{13}$C = $^{13}$C bond 
length of $r \approx$ 0.135nm. \cite{Ref_34}}
\end{figure} 

	The angular dependence of the splitting width, $d$, for each doublet can be also explained by the orthorhombic symmetry [Figs.~\ref{Fig_3}(a) and \ref{Fig_3}(b)]. 
	The solid and dashed curves represent the calculated angular dependence of $d$ expressed as $d=(3/2r^{3})\gamma_{\textrm{n}}^{2}\hbar(1-3\cos^{2}\Theta)$, 
\cite{Ref_29_Kawamoto} where $\gamma_{\textrm{n}}/2\pi$ = 10.7054MHz/T is the gyromagnetic ratio 
of $^{13}$C nucleus, 2$\pi\hbar$ is the Planck constant, $r$ denotes the distance 
between the adjacent $^{13}$C nuclei, and $\Theta$ is the angle between $H$ and the molecular 
$x$ axis [see Fig.~\ref{Fig_1}(c)]. 
	We used the orthorhombic sublattice structure of BEDT-TTFs reported by Kobayashi \textit{et al}. \cite{Ref_33_Kobayashi} 
and the optimal $^{13}$C = $^{13}$C bond length $r \approx$ 0.135nm. \cite{Ref_34} 
	The calculated curves are in good agreement with the observed splitting width.

\begin{figure}
  \includegraphics[width=8.6cm]{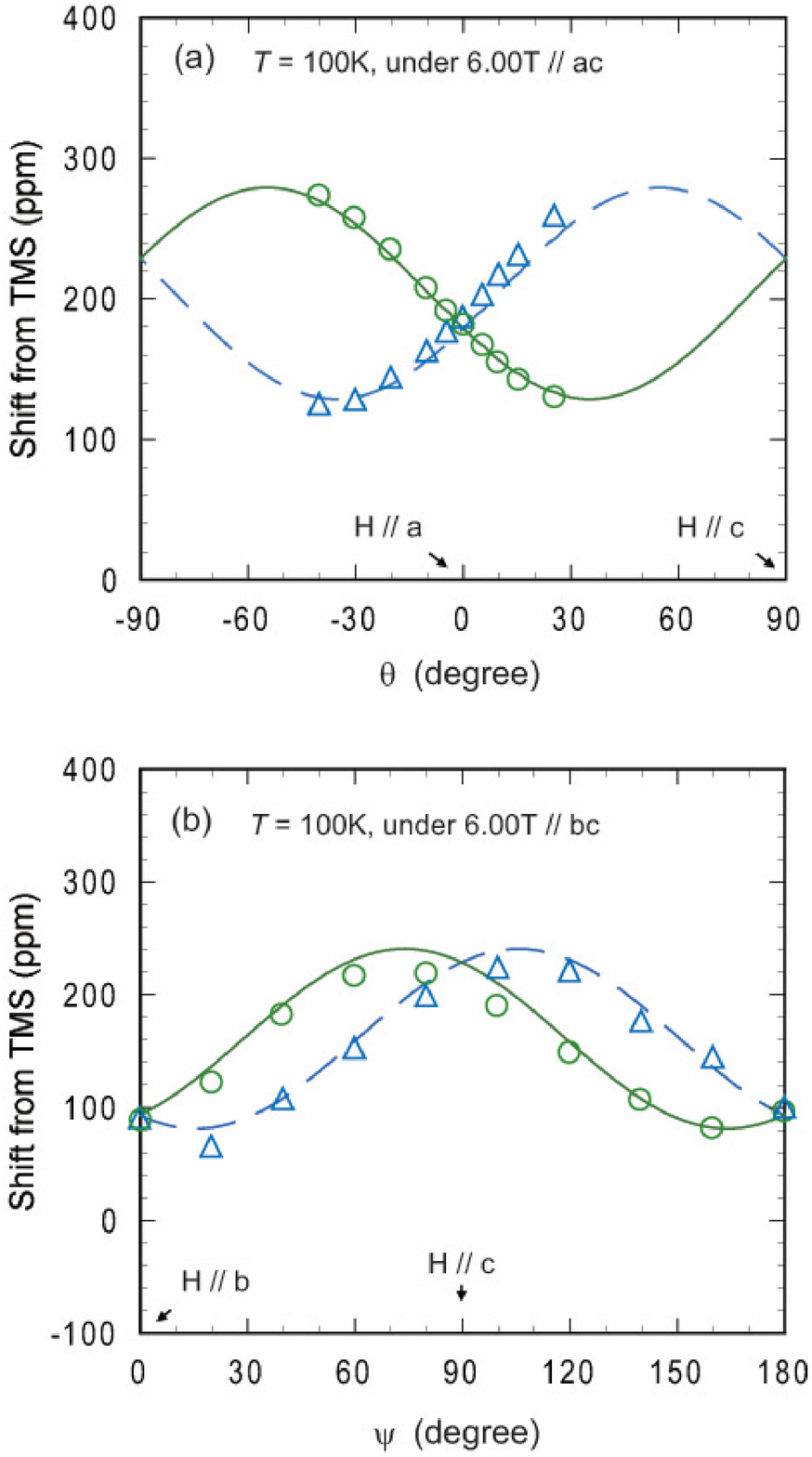}
  \caption{\label{Fig_4} (Color online) Angular dependence of the $^{13}$C-NMR shift $\delta$ 
(central line shift of a Pake doublet) at 
(a) $H//ac$ (100K) and (b) $H//bc$ (100K) for the two unequal molecules in the unit cell (symbols; see Fig.~\ref{Fig_1}). 
	The same symbols are used as in Figs.~\ref{Fig_2} and \ref{Fig_3}. 
	Curves stand for the least square fits to the data calculated from the orthorhombic structure \cite{Ref_33_Kobayashi} 
and the hyperfine-shift tensor given in the text.}
\end{figure}

	Figures~\ref{Fig_4}(a) and \ref{Fig_4}(b) depict the angular dependence of the $^{13}$C-NMR 
shift at 100K given by the midpoint of each Pake doublet under $H//ac$ and $H//bc$, 
respectively (the same symbols as in Figs.~\ref{Fig_2} and \ref{Fig_3} are used). 
	We performed a least square fit of sinusoidal curves to these data sets 
simultaneously with the three principal values of 
the $^{13}$C hyperfine-shift tensor as fitting parameters on the basis of 
the orthorhombic structure. \cite{Ref_33_Kobayashi} 
	The resulting fitting curves are shown by solid and dashed lines 
in Figs.~\ref{Fig_4}(a) and \ref{Fig_4}(b), where the principal values of the $^{13}$C hyperfine-shift 
tensor are uniquely determined as 
($\delta_{xx}, \delta_{yy}, \delta_{zz}$) = (67, 134, 300) (in ppm). 
	Again, the agreement between the calculated curves and the experimental data are well.

	From all these arguments, it is thus concluded that all of the BEDT-TTF molecules in $\theta$-I$_{3}$ 
are electronically equivalent within experimental accuracy, indicating that the local-site symmetry at BEDT-TTFs can be practically seen as orthorhombic even though the crystal structure is monoclinic. 
This suggests that the three dimensional network of the triiodine sublattice, 
which gives rise to the monoclinic symmetry of $\theta$-I$_{3}$,\cite{Ref_08_Kobayashi} 
contributes little to the reduction of the electronic and structural symmetries in the BEDT-TTF layers, 
and hence, when we look into the conducting layer solely, the BEDT-TTF sublattice is to be treated approximately as orthorhombic. 
This is in a good agreement with the quantum oscillation \cite{Ref_09_Kajita, Ref_10_Klepper, 
Ref_11_Tokumoto, Ref_12_Tamura, Ref_13_Terashima, Ref_14_Salameh} and 
optical reflectance \cite{Ref_15_Tamura, Ref_16_Oshima} experiments 
both of which point to a large Fermi surface expected in the orthorhombic symmetry. \cite{Ref_08_Kobayashi}

	It is also noteworthy to mention that each line in the NMR spectra is, in the whole temperature range,
much narrower ($\sim$ 1kHz) than the cases in $\theta$-RbZn and $\theta$-CsZn, 
where NMR lines show a splitting or broadening over several kHz or more in association with 
the charge ordering or its glassy freezing, respectively. \cite{Ref_05_Miyagawa, Ref_19_Chiba}
	The present spectral feature clearly indicates that $\theta$-I$_{3}$ at ambient pressure 
is free from these instabilities of the charge organization. 
	This is consistent with the sharp line width observed in 
Raman scattering experiments. \cite{Ref_35_Wojciechowski}

	As we mentioned in Section~\ref{intro}, the metallic phase of $\theta$-I$_{3}$ is suggested to turn into a 
massless Dirac fermion system with a linear energy-momentum dispersion above $\sim$ 0.5GPa, 
\cite{Ref_20_Tamura, Ref_21_Tajima} as confirmed by our previous $^{13}$C-NMR shift and 
relaxation-rate $1/T_{1}$ measurements at 0.8GPa. \cite{Ref_22_Miyagawa}
	Noticeably, NMR spectra in the high-pressure phase exhibit more than eight lines. 
	The qualitative difference in the spectral feature below and above $\sim$ 0.5GPa evidences that 
the transition at $\sim$ 0.5GPa accompanies a structural phase transition with a symmetry reduction; 
thereby $\theta$-I$_{3}$ is very probably transformed into a new structure similar to 
$\alpha$-I$_{3}$ that accommodates the massless Dirac fermions.

\subsection{Knight shift, nuclear spin-lattice relaxation rate, and electron-correlation effect}
\label{3b}
	Next, we turn our attentions to the local electronic structures in $\theta$-I$_{3}$ 
at ambient pressure. 
	We measured the temperature dependence of $^{13}$C-NMR spectra, 
which gives Knight shift, $K$, and nuclear spin-lattice relaxation rate, $1/T_{1}$. 

\begin{figure}
  \includegraphics[width=16cm]{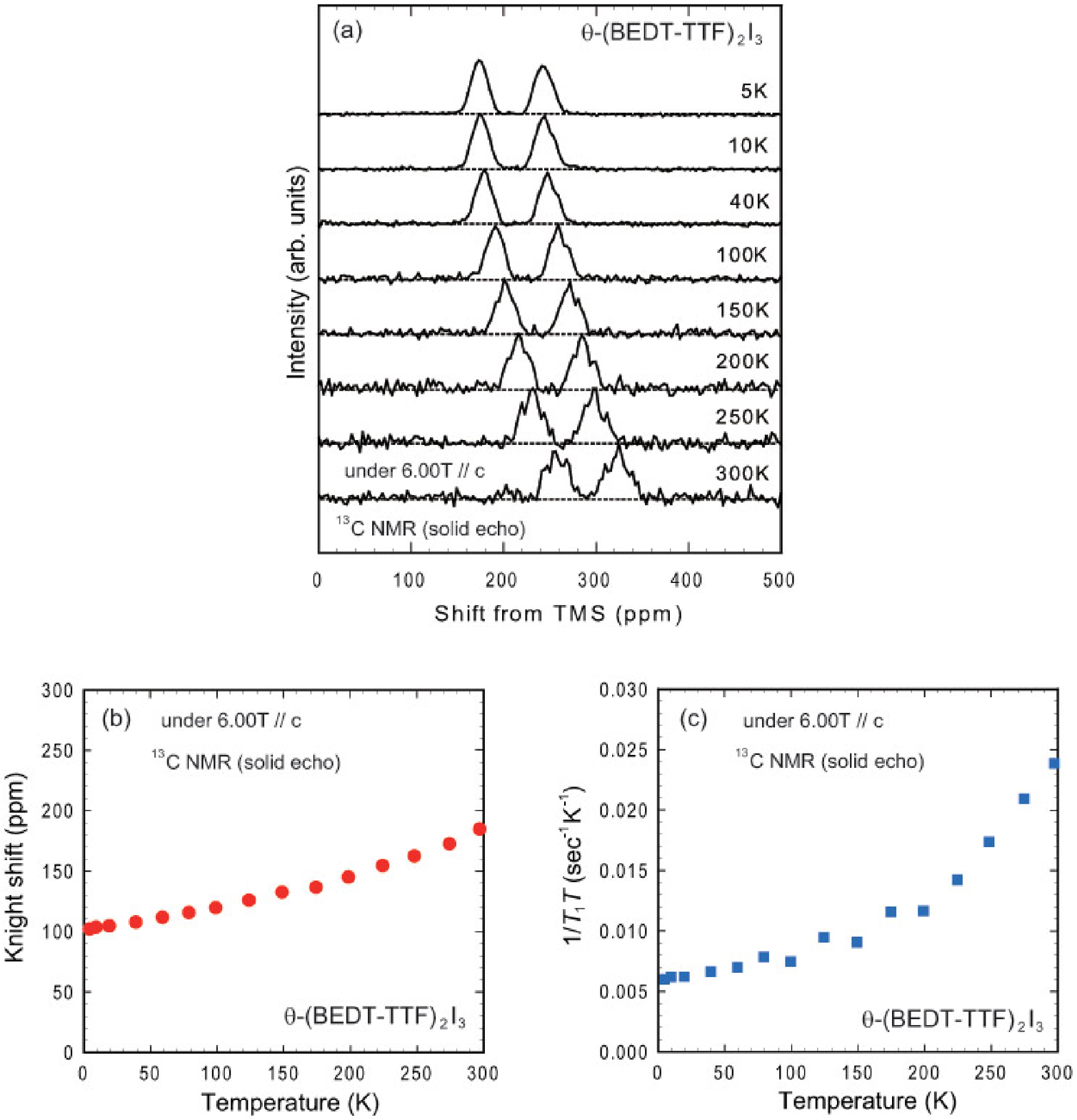}
  \caption{\label{Fig_5} (Color online) (a) Temperature dependence of the $^{13}$C-NMR 
spectra under $H//c$. 
(b), (c): Temperature dependence of the Knight shift $K$ (= $\delta - \sigma$; 
with the central NMR line shift $\delta$ and the temperature-independent chemical shift 
$\sigma \approx$ 106ppm; see the text) and the nuclear spin-lattice relxation rate divided by temperature $1/T_{1}T$ 
 under the same field orientation.}
\end{figure}

	Figure~\ref{Fig_5}(a) shows the temperature dependence of the $^{13}$C-NMR spectra 
under an external field $H$ of 6.00T applied parallel to the $c$ axis in the conduction layer. 
	All molecules become equivalent under this field orientation, 
so only a single Pake doublet is observed.

	The spectra show moderate temperature dependence with decreasing temperature. 
	The NMR shift, $\delta$, defined as the midpoint of the doublet, is composed of 
the Knight shift $K$ and the chemical shift, $\sigma$, terms, i.e., 
$\delta = K + \sigma$. 
The Knight shift $K$ is related to the electron spin susceptibility $\chi_{\textrm{s}}$ via 
$K = a\chi_{\textrm{s}}/2\mu_{\textrm{B}}N_{\textrm{A}}$, where $a$ is the 
hyperfine-coupling constant, $\mu_{\textrm{B}}$ is the Bohr magneton, and $N_{\textrm{A}}$ is 
Avogadoro's number. 
	The chemical shift $\sigma$ originates from the orbital motion of electrons in 
a BEDT-TTF molecule and is temperature independent. 
	In the present field configuration, it is estimated as $\sigma \approx$ 106 ppm from the 
chemical-shift tensors reported by Kawai \textit{et al}. \cite{Ref_36_Kawai, Ref_37} and 
the orthorhombic structure. \cite{Ref_33_Kobayashi} 
	By subtracting $\sigma$ from the NMR shift $\delta$, we evaluated the Knight shift 
$K$ (= $\delta - \sigma$) as shown in Fig.~\ref{Fig_5}(b). 
	Knight shift $K$ decreases monotonically with temperatures with a slight concave around 200K, 
approaching a value of $K \approx$ 100ppm in the low-temperature limit. 

	The temperature dependence of Knight shift $K$ is more prominent 
than that of the spin susceptibility $\chi_{\textrm{s}}$ determined 
by magnetization measurements earlier. \cite{Ref_14_Salameh}
It is not clear what causes the difference in the temperature 
dependence of $\chi_{\textrm{s}}$ and $K$ in $\theta$-I$_{3}$.
If this is attributable to sample dependence, one possibility is 
the randomness in the position of triiodine molecules as reported by 
x-ray diffraction experiments;\cite{Ref_08_Kobayashi} 
I$_{3}$ molecules have two possible positions along the $c$-axial direction 
that are mutually shifted by $c$/2, where $c$ (= 0.496nm) is 
the lattice constant along the $c$ axis.\cite{Ref_08_Kobayashi} 
This randomness gives a disorder in the triiodine positions, 
which may affect the electronic properties. 
However, if the I$_{3}$ disorder had a large influence on the electronic states, 
the NMR spectra would exhibit eight lines 
as they should reflect the whole crystal structure 
and the monoclinic symmetry of the crystal does produce eight lines 
(; see Sec.~\ref{3a}). 
Obviously, this is not the case as seen in our previous arguments in Sec.~\ref{3a}.
Furthermore, the observed line width is so sharp ($\sim$ 1kHz) 
as to be resolution limited. 
These facts indicate that the disorder in the I$_{3}$ sublattice, if any, 
is not influential on the electronic states in the BEDT-TTF sublattice 
at least for the present sample.

	In Fig.~\ref{Fig_5}(c), we show the temperature dependence of 
nuclear spin-lattice relaxation rate $1/T_{1}$ divided by temperature, $1/T_{1}T$, 
under the same field orientation ($H//c$). 
	1/$T_{1}T$ exhibits a monotonic decrease with decreasing temperature. 
	Compared with the Knight shift $K$ shown in Fig.~\ref{Fig_5}(b), $1/T_{1}T$ shows relatively 
large temperature dependence, which seems to get steeper above ca. 200K. 
	In a metallic system, $1/T_{1}$ is proportional to the scattering rate of conducting electrons 
by nuclear spins near the Fermi level $E_{\textrm{F}}$ and is known, in case of an isotropic hyperfine 
interaction characterized by the coupling constant $a$, to follow the Korringa relation 
\cite{Ref_38_Slichiter} --- $1/T_{1}T = 4\pi k_{\textrm{B}}/\hbar (\gamma_{\textrm{n}}/\gamma_{\textrm{e}})^{2}a^{2}(\mu_{\textrm{B}}/2N_{\textrm{A}})^{2}<D(E_{\textrm{F}})^{2}>_{T}$ --- 
where $D(E_{\textrm{F}})$ is the electronic density of states at $E_{\textrm{F}}$, 
$\gamma_{\textrm{e}}$ is the gyromagnetic ratio of electron, and $<x>_{T}$  stands 
for the thermal average of the quantity $x$. 
	In a conventional metal without electron-electron correlations, 
the following relation holds between Knight shift $K$ and relaxation rate $1/T_{1}$; 
$(1/T_{1}T)K^{-2}(\hbar/4\pi k_{\textrm{B}})(\gamma_{\textrm{e}}/\gamma_{\textrm{n}})^{2}=1$. 
	In (BEDT-TTF)$_{2}$X compounds, however, the hyperfine interaction is strongly anisotropic 
at the $^{13}$C positions because of a significant contribution of dipolar interactions from 
the $p_{\textrm{z}}$ orbitals. 
	This effect introduces a geometrical form factor in the above relation, namely, 
$(1/T_{1}T)K^{-2}(\hbar/4\pi k_{\textrm{B}})(\gamma_{\textrm{e}}/\gamma_{\textrm{n}})^{2}= 
\beta(\zeta , \eta)$ \cite{Ref_29_Kawamoto, Ref_32_Hirata} with 

\begin{equation}
\beta(\zeta , \eta) = 
\frac{(a^{xx}/a^{zz})^{2}(\sin^{2}\eta +\cos^{2}\zeta\cos^{2}\eta + (a^{yy}/a^{zz})^2(\cos^{2}\eta +\cos^{2}\zeta\sin^{2}\eta) + \sin^{2}\zeta }{2[(a^{xx}/a^{zz})\sin^{2}\zeta\cos^{2}\eta + (a^{yy}/a^{zz})\sin^{2}\zeta\sin^{2}\eta +\cos^{2}\zeta]^{2}},
\end{equation}
where ($a^{xx}, a^{yy}, a^{zz}$) are the principal values of the $^{13}$C-hyperfine-coupling tensor, 
$\zeta$ is the angle between the external field $H$ and the $z$-principal axis shown 
in Fig.~\ref{Fig_1}(c), and $\eta$ is the polar angle measured from the $x$-principal axis 
in the $xy$ plane of Fig.~\ref{Fig_1}(c). $\beta (\zeta, \eta)$ is dependent on the orientation of 
the applied field $H$ relative to the molecular axes. 
	Furthermore, the effect of the electron correlations is incorporated into the form by the 
introduction of a factor, ${\cal K}$, called the Korringa ratio which measures 
the degree of spin fluctuations quantitatively, \cite{Ref_29_Kawamoto} i.e., 
$(1/T_{1}T)K^{-2}(\hbar/4\pi k_{\textrm{B}})(\gamma_{\textrm{e}}/\gamma_{\textrm{n}})^{2}= 
\beta(\zeta , \eta){\cal K}$ . 
${\cal K}$ is unity for free electrons, 
while it becomes larger (smaller) than unity if there are antiferromagnetic (ferromagnetic) 
spin fluctuations. \cite{Ref_39_Moriya} 
	Using the hyperfine-shift tensor determined in Section \ref{3a}, 
the susceptibility $\chi_{\textrm{s}}$ at 100K reported by Salameh \textit{et al}. 
\cite{Ref_14_Salameh} ($\chi_{\textrm{s}} = 4.9 \times 10^{-4}$ emu/mol f.u.), 
and the chemical-shift tensors determined by Kawai \textit{et al}., \cite{Ref_36_Kawai, Ref_37}
we estimated the principal values of the $^{13}$C-hyperfine-coupling tensor as (-1.1, -0.9, 5.6) 
(in kOe/$\mu_{\textrm{B}}$). 
	Then, $\beta(\zeta, \eta)$ is evaluated as $\beta(\zeta, \eta) \approx$ 0.97 with $(\zeta, \eta) = 
(139^{\circ}, 85^{\circ})$ under the current field configuration $(H//c)$.

\begin{figure}
  \includegraphics[width=8.6cm]{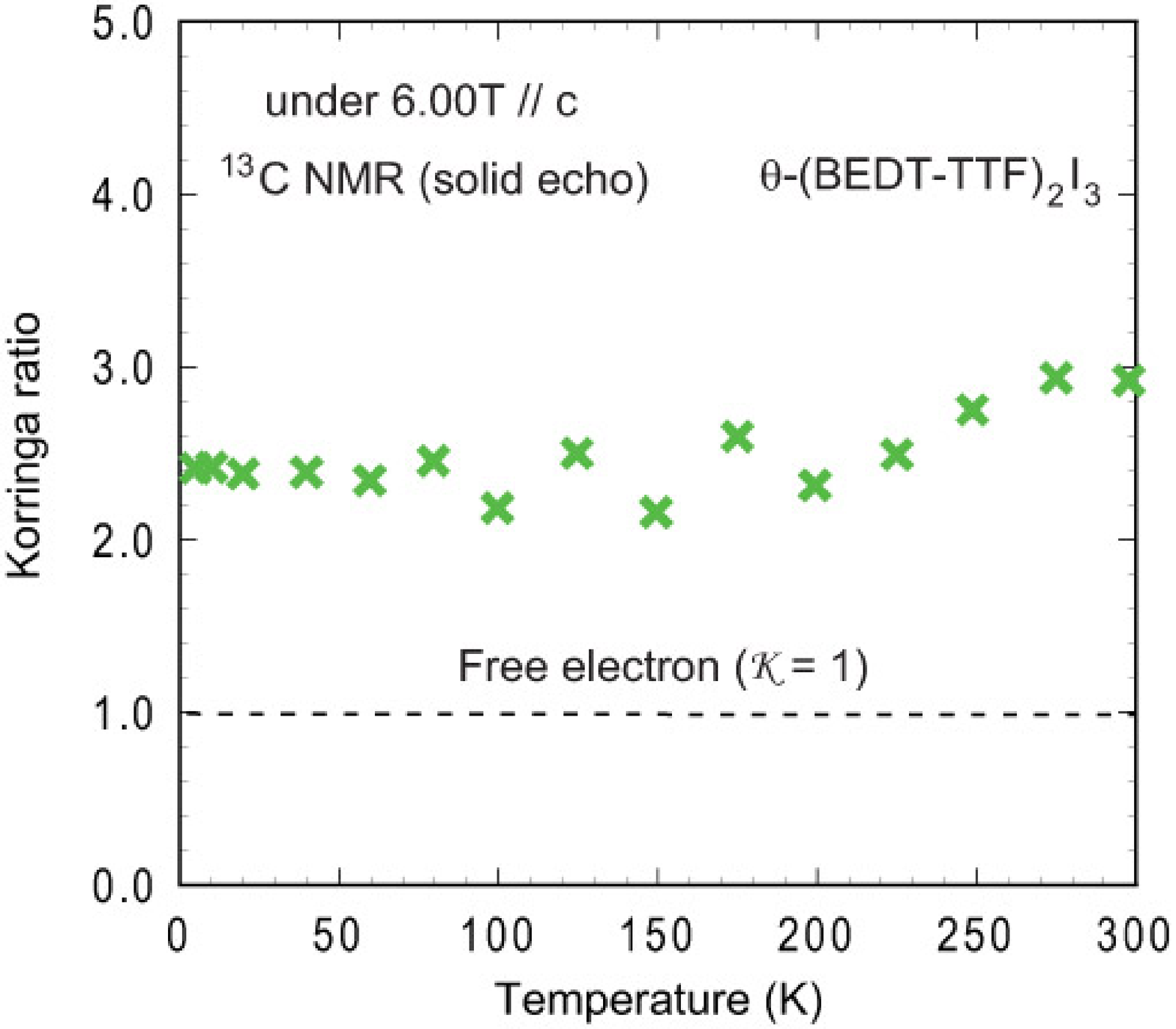}
  \caption{\label{Fig_6} (Color online) Temperature dependence of the $^{13}$C-NMR Korringa ratio, 
${\cal K}\propto 1/(T_{1}TK^{2})$, under $H//c$. 
The dashed line stands for the free electron's case (${\cal K}$ = 1; see the text).}
\end{figure}

	With the use of Knight shift $K$ [Fig.~\ref{Fig_5}(b)] and $1/T_{1}T$ [Fig.~\ref{Fig_5}(c)], 
and the value of $\beta(\zeta, \eta)$ determined above, the temperature dependence of the Korringa ratio 
${\cal K}$ is obtained as shown in Fig.~\ref{Fig_6}. 
	There is no strong temperature dependence in ${\cal K}$ 
with ${\cal K}$ = 2.4 $\pm$ 0.2 up to $\sim$ 200K, which tends to increase into a value of ${\cal K} \sim$ 3.0 
around room temperature. 
	The value greater than unity points to the presence of antiferromagnetic spin fluctuations 
but is appreciably smaller than the typical values in strongly correlated electron systems realized 
in similar organic conductors; 
for instance, the 2D metal with strong electron correlations, 
$\kappa$-(BEDT-TTF)$_{2}$Cu[N(CN)$_{2}$]Br and $\kappa$-(BEDT-TTF)$_{2}$Cu(NCS)$_{2}$, 
\cite{Ref_40_Mayaffre, Ref_41_Soto, Ref_42_Itaya} locating in the vicinity of Mott transition, 
\cite{Ref_02_Miyagawa} show the Korringa ratio of ${\cal K} \sim$ 8 at ambient pressure. 
	The present value of ${\cal K}$ hence indicates that electrons in $\theta$-I$_{3}$ are 
moderately correlated. 
	This is consistent with the experiments of quantum oscillations 
\cite{Ref_09_Kajita, Ref_10_Klepper, Ref_11_Tokumoto, Ref_12_Tamura, Ref_13_Terashima, 
Ref_14_Salameh} giving an effective mass which is approximately a half of the $\kappa$-type 
salts'.\cite{Ref_add_2_Caulfield} 

	The overall results show no signature of charge ordering unlike most of $\theta$-type 
compounds. 
	Most importantly, however, it should be noted that the steep increase in $1/T_{1}T$ above $\sim$ 200K 
[Fig.~\ref{Fig_5}(c)] is a feature not seen in other metallic salts \cite{Ref_43_Kanoda} and implies 
an enhancement of spin scatterings. 
	The feature that the enhancement is less prominent in ${\cal K}$ (Fig.~\ref{Fig_6}) 
suggests that the spin scatterings are distributing over a wide range in $\bm{k}$-space 
and are irrelevant to antiferromagnetic instabilities. 
	Recently, Cano-Cortes and coworkers \cite{Ref_44_Cano} theoretically investigated the 
extended Hubbard model with a modeled $\theta$-type structure. 
	Among the consequences is the possible quantum phase transition of charge order with respect 
to the strength of electron correlations. \cite{Ref_44_Cano, Ref_45_Dressel} 
	In the metallic side in the phase diagram, the system is predicted to exhibit a crossover from 
a Fermi liquid to an incoherent bad metal with quantum critical charge fluctuations 
at a finite temperature $T^{\ast}$, which depends on the distance from the quantum critical point 
in the phase diagram. 
	In the incoherent state above $T^{\ast}$, the spin lifetime is expected to be shortened 
without preferential scattering vector. 
	The turnabout behavior in $1/T_{1}T$ around 200K can be a signature of this crossover. 
	Assuming that the largest transfer integral, $t_{a}$, is $t_{a} \sim$ 80meV in $\theta$-I$_{3}$
as deduced from the analysis of optical data, \cite{Ref_15_Tamura} $T^{\ast} \sim$ 200K corresponds to 
a reduced temperature of $T^{\ast}/ t_{a} \sim$ 0.2, which is in a reasonable range 
in the prediction. \cite{Ref_44_Cano}
	A related feature is seen in the optical conductivity measurements 
by Takenaka \textit{et al}. \cite{Ref_46_Takenaka} who observed a rapid loss of electronic 
coherence with increasing temperature. 
	These experimental and theoretical results suggest that $\theta$-I$_{3}$ is situated not 
so far away from the charge-ordering point.
	In case that spin fluctuations develop toward an antiferromagnetic ordering, 
the Korringa ratio ${\cal K}$ should be temperature sensitive, reflecting the preferential 
development of antiferromagnetic fluctuations with a well-defined wave vector $\bm{Q}$ 
relative to the $\bm{Q}$ = 0 (ferromagnetic) fluctuation. 
	If the charge sector is also fluctuating, however, spin fluctuations should 
be spread in $\bm{Q}$ and may be only moderately reflected in ${\cal K}$.
\section{CONCLUSION}
	We performed $^{13}$C-NMR measurements for a single crystal of 
$\theta$-(BEDT-TTF)$_{2}$I$_{3}$ at ambient pressure to probe microscopically the nature of 
the metallic state, which is a rare case in the $\theta$-type family of compounds. 
	The orientation dependence of the NMR spectra revealed that the lattice symmetry can be practically 
understood by the orthorhombic symmetry with a homogeneous molecular arrangement. 
	The NMR shift and relaxation rate $1/T_{1}$ measurements showed the Korringa relation up 
to room temperature with the weakly temperature-dependent Korringa ratio ${\cal K}$ (= 2 -- 3), 
which indicates that $\theta$-(BEDT-TTF)$_{2}$I$_{3}$ is in a moderately correlated regime. 
	However, the enhancement of relaxation rate, observed above $\sim$ 200K, 
is a possible signature of the quantum critical charge fluctuations as suggested 
by a recent theoretical study.

\section{ACKNOWLEDGEMENTS}
	The authors thank H. Kobayashi for informing us of the structural data before publication
and for helpful and kind discussions. 
	This work is supported by MEXT Grant-in-Aids for Scientific Research on Innovative Area 
(New Frontier of Materials Science Opened by Molecular Degrees of Freedom; Grants No. 20110002 
and No. 21110519), JSPS Grant-in-Aids for Scientific Research (A) (Grant No. 20244055) and (C) 
(Grant No. 20540346), and MEXT Global COE Program at University of Tokyo 
(Global Center of Excellence for the Physical Sciences Frontier; Grant No. G04).

\end{document}